\documentclass[aps,prapplied,reprint,showkeys,superscriptaddress,floatfix]{revtex4-1}
\usepackage{amsfonts}
\usepackage{amssymb}
\usepackage{amsmath}
\usepackage{hyperref}
\usepackage{graphicx}
\usepackage{bookmark}
\begin{document}
\title{Graphene Aerogels for Ultrabroadband Thermoacoustics}
\author{Francesco De Nicola}
\email[E-mail: ]{francesco.denicola@iit.it}
\affiliation{Graphene Labs, Istituto Italiano di Tecnologia, Via Morego 30, 16163 Genova, Italy}
\affiliation{Department of Physics, University of Rome La Sapienza, P.le A. Moro 5, 00185 Rome, Italy}
\author{Stefano Sarti}
\affiliation{Department of Physics, University of Rome La Sapienza, P.le A. Moro 5, 00185 Rome, Italy}
\author{Bing Lu}
\affiliation{Beijing Institute of Technology, 100081 Beijing, China}
\author{Liangti Qu}
\affiliation{Beijing Institute of Technology, 100081 Beijing, China}
\author{Zhipan Zhang}
\affiliation{Beijing Institute of Technology, 100081 Beijing, China}
\author{Augusto Marcelli}
\affiliation{INFN-Laboratori Nazionali di Frascati, Via Enrico Fermi 40, 00044 Frascati (RM), Italy}
\affiliation{RICMASS, Rome International Center for Materials Science Superstripes, Via dei Sabelli 119A, 00185 Rome, Italy}
\author{Stefano Lupi}
\affiliation{Graphene Labs, Istituto Italiano di Tecnologia, Via Morego 30, 16163 Genova, Italy}
\affiliation{Department of Physics, University of Rome La Sapienza, P.le A. Moro 5, 00185 Rome, Italy}
\date{\today}
\begin{abstract}
Sound is usually generated in a medium by an electromechanical vibrating structure. The geometrical size and inertia of the structure set the frequency cutoff in the sound-transduction mechanism and, often, different vibrating structures are necessary to cover the whole range from infrasound to ultrasound. An alternative mechanism without any physical movement of the emitter is the thermoacoustic effect, where sound is produced by Joule heating in a conductive material. Here we show that a single thermoacoustic transducer based on a graphene aerogel can emit ultrabroadband sound from infrasound (1 Hz) to ultrasound (20 MHz), with no harmonic distortion. Since conventional acoustic transducers are frequency band limited due to their transduction mechanism, ultrabroadband graphene aerogels may offer a valid alternative to conventional hi-fi loudspeakers, and infrasound and ultrasound transducers.
\end{abstract}
\maketitle
\section{Introduction}
\label{sec:intro}
\indent Modern wireless communication is based on transmitting and receiving electromagnetic waves that span a wide frequency range, from hertz to terahertz, providing a large spectral interval for high data transfer rates. There are drawbacks in electromagnetic communication, though, including high extinction coefficient for electrically conductive materials and the large size of antennas \cite{Zhou2005}.\\
\indent However, animals have effectively used acoustic waves for short-range communication for millions of years \cite{Tian2014a}. Acoustic wave-based communication, although embodying a reduced spectral band due to the geometrical size and inertia of the vibrating structure that set the frequency cut-off in the sound transduction mechanism \cite{Beranek1993}, can overcome some of the difficulties in electromagnetic wave-based communication and complement existing wireless technologies. For example, acoustic waves propagate well in conductive materials, thus have been explored for underwater communication (SONAR) \cite{Kilfoyle2000}. Animals, such as rodents, bats, and cetaceans, are known to communicate and move (eco-localization) effectively by ultrasound waves (20-300 kHz) \cite{Tian2014a}. In land-based acoustic wave communication, the audible band (20 Hz-20 kHz) \cite{Beranek1993} is often occupied by human conversations and acoustic loudspeakers, whereas the infrasonic band ($<20$ Hz) can be disturbed by moving vehicles and building construction \cite{Everest2001}. Although infrasonic waves are usually annoying for humans and animals, brain waves (1-600 Hz) take place in this spectral band \cite{Schnitzler2005}. Despite having a wide frequency range and often free of disturbance, the ultrasonic band is rarely exploited for high data rate communication purposes. One possible reason for this is the lack of wide bandwidth ultrasonic emitters and receivers. Conventional ultrasound transducers \cite{Hedrick2005} exploit the piezoelectric effect, thus they only operate near their resonance frequencies due to their membrane vibrations. Another typical issue for acoustic transducers is the frequency-dependent impedance of the materials employed, which contributes to limit their spectral band\cite{Beranek1993}. As a result, there is no broadband acoustic transducer to date that covers from infrasound to ultrasound, preventing the broadband or multiband transmission and detection of signals for communications.\\
\indent An alternative mechanism for sound transduction is the thermoacoustic (TA) effect\cite{Giorgianni2018,Guiraud2019,Tian2014a,DeNicola2019,Kim2016,Fei2015,Aliev2015,Tian2011,Tian2012,Suk2012,Xu2013,Tian2014,Tian2015,Kim2016a,Tao2017,Heath2017}, whereas sound is emitted without moving part by Joule heating when an electric current flows in a conductive material. The root-mean-square sound pressure amplitude $p_{rms}$ of a TA wave can be derived by a general analytical solution of the TA model \cite{Vesterinen2010,Hu2010}, as follow
\begin{equation}
p_{rms}=\frac{Rv_{g}q_{0}}{\sqrt{2}r_{0}C_{p,g}T_{g}}\frac{e_{g}}{M(f)e_{s}+e_{g}}\mathcal{D}(\theta,\phi),
\label{eq:TA}
\end{equation}
where $R=Sf/v_{g}$ is the Rayleigh factor with $S$ the emitting surface area of the material, $f$ the sound frequency, and $v_{g}$ the speed of sound in the medium, $q_{0}=Q_{0}/S$ is the oscillating power density at frequency $f$ dissipated by the material with $Q_{0}=|V_{t}|^{2}/|Z|$ the amplitude of the oscillating component of the electric power, being $V(t)=V_{0}\sin{(2\pi ft)}$ the applied AC voltage and $Z$ the material electrical impedance, $r_{0}$ is the far-field distance from the sound source, $e_{i}=\sqrt{k_{i}\rho_{i}C_{p,i}}$ is the thermal effusivity of the material $(i=s)$ and the medium $(i=g)$ with $k_{i}$ the thermal conductivity, $\rho_{i}$ the mass density, and $C_{p,i}$ the specific heat capacity at constant pressure, $T_{g}$ is the medium temperature, $M(f)\approx1$ is a frequency-dependent factor, and $\mathcal{D}(\theta,\phi)$ is the far-field directivity. From the above equation is evident that the TA effect is not limited in frequency, potentially enabling broadband TA transducers.\\
\indent Here, we introduce a transducer based on a graphene aerogel able to generate ultrabroadband sound from 1 Hz up to 20 MHz by TA effect. The sound emitted has no observable harmonic distortion in the whole range of frequency investigated and the aerogels act as an omni-directional point source up to 20 kHz. Furthermore, we demonstrate that, among the other TA transducers \cite{Aliev2015}, graphene has the advantage to have a constant electrical impedance from DC up to the MHz range, allowing no frequency cut-off or specific resonances, unlike in conventional acoustic transducers \cite{Beranek1993}. The present research represents a breakthrough for audio consumer technologies, underwater communication \cite{Kilfoyle2000}, transducers for zoological \cite{Tian2014a} and biomimetic \cite{Kihara2006} applications, devices for medical diagnostics \cite{Szabo2004} and food quality control \cite{Awad2012}.
\section{Experimental}
\label{sec:exp}
\subsection{Graphene aerogel fabrication}
\label{sec:graphene}
\indent The aerogels were obtained by a graphene oxide dispersion (60 mL, 4-9 mg/mL) and 7 mL of Tween 80 (150 mg/mL) ethanol solution mixed at 2500 rpm for 5 min. The mixture was exposed to liquid nitrogen for 30 min. After freeze-drying for 72 h, the mixture was heated at $300^{\circ}$C for 2 h in an Ar atmosphere.\\
\indent The graphene aerogels have a cylindrical shape with a diameter $d=4$ cm, a thickness $L=6.5$ cm, and a mass $m=500$ mg, therefore a mass density $\rho_{s}=6$ kg/m$^{3}$. From the scanning electron microscopy (SEM) micrograph of a representative sample in Figure \ref{fig:Figure1}a, it is possible to observe a porous ($\approx80\%$) random network made of graphene flakes constituting the aerogel surface.
\subsection{Acoustic characterization}
\label{sec:acoustics}
\begin{figure}[ht!]
\centering
\includegraphics[width=0.45\textwidth]{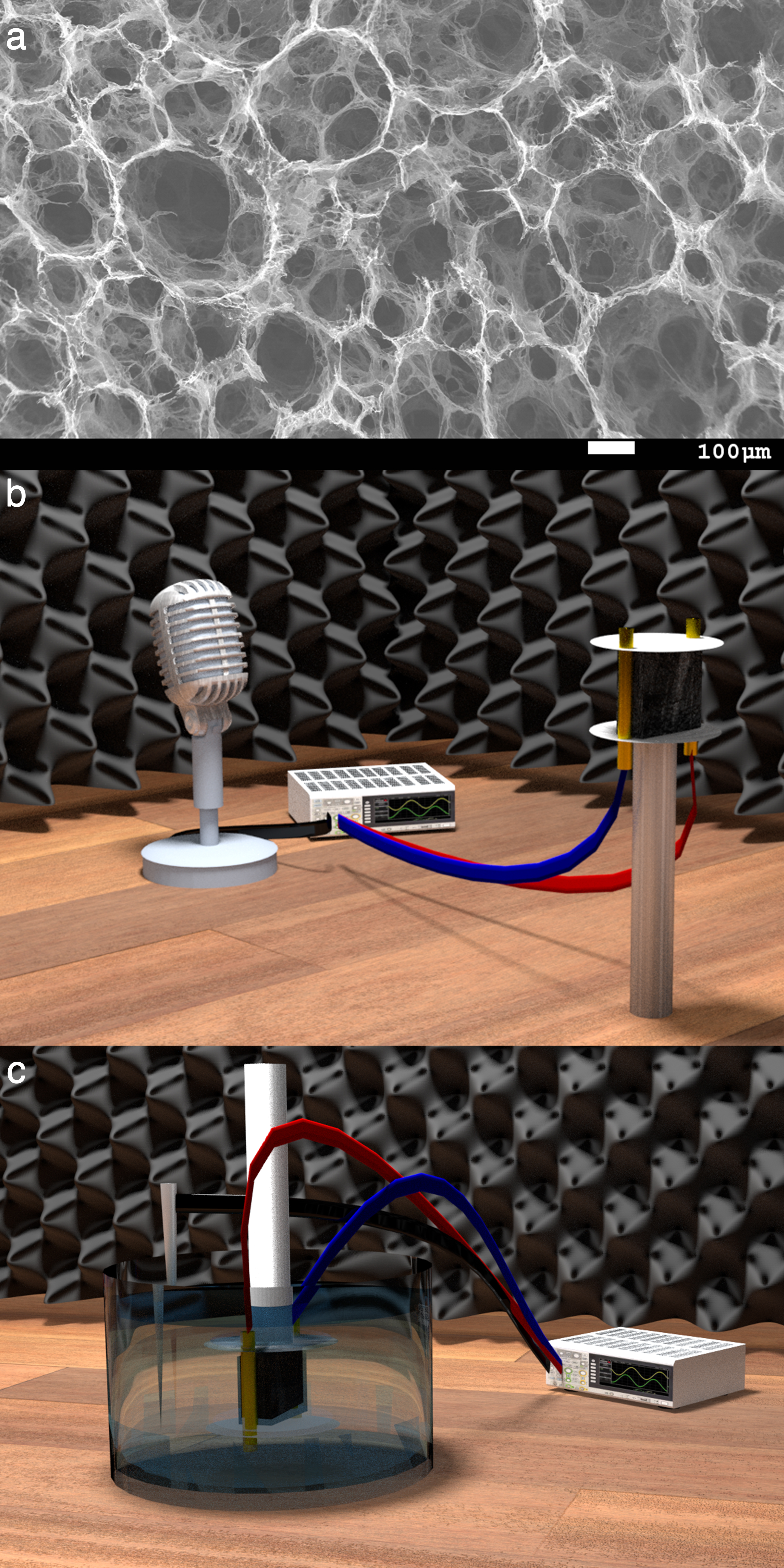}
\caption{Graphene aerogel TA transducers. \textbf{a}, Representative SEM micrograph of a graphene aerogel. \textbf{b}, Conceptual scheme of the experiment in air. An AC voltage is applied at a graphene aerogel sample. The sound emitted by the aerogel is recorded by a microphone connected to a sound card. \textbf{c}, Conceptual scheme of the experiment in liquid. An AC voltage is applied at a sample placed in a beaker filled with water. The sound emitted by the aerogel is detected by a needle hydrophone immersed in water and placed at a given angle with respect the sound emitting surface of the aerogel. The hydrophone is pre-amplified and connected to a DC coupler and an oscilloscope.}
\label{fig:Figure1}
\end{figure}

\indent The experimental setup was placed in a sound-proof room ($3\times2\times2$ m) in order to acoustically insulate the experimental apparatus from environmental noise and to reduce internal sound reflections. The audio signal flow was calibrated in an anechoic chamber at the National Institute for Insurance against Accidents at Work (INAIL).\\
\indent We measured the acoustic emission of graphene aerogels both in air and water, in order to avoid the ultrasound ($f>20$ kHz) absorption by air. The scheme of the experimental setups in air and water is reported in Figure \ref{fig:Figure1} (b, c) respectively.\\
\begin{figure*}[ht!]
\centering
\includegraphics[width=1\textwidth]{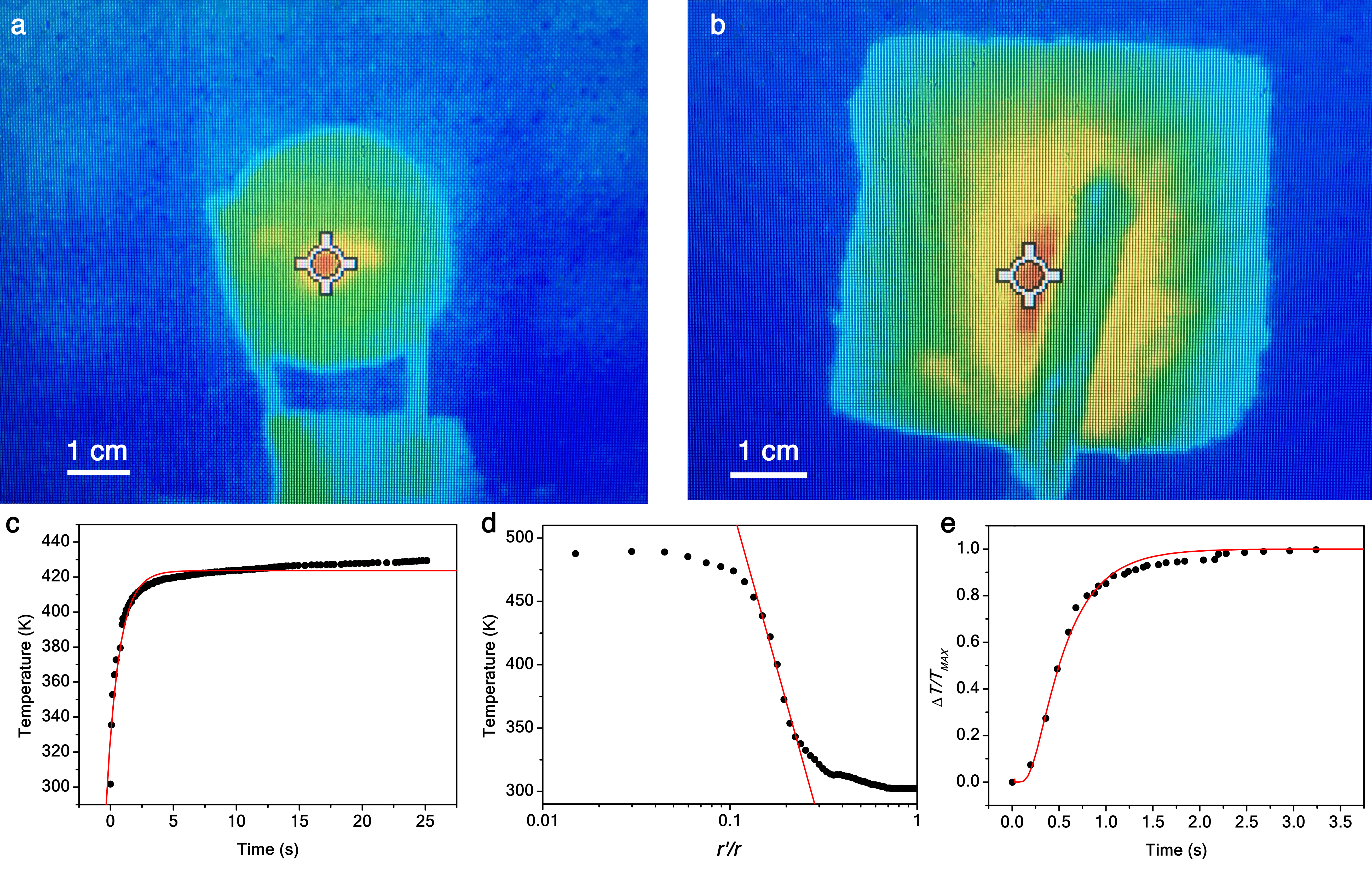}
\caption{\textbf{a}, Thermal profile image of the heated sample base surface. \textbf{b}, Thermal profile image of the heated sample lateral surface. \textbf{c}, Temperature rise as a function of time on axis of the graphene aerogel. Data is fit by Equation 2 (red solid line). \textbf{d}, Radial thermal profile of the graphene aerogel illuminated surface as a function of the relative coordinate. Data is fit by Equation 3 (red solid line). \textbf{e}, Temperature rise as a function of time, at distance $L$ from the graphene aerogel heated surface. Data is fit by Equation 4 (red solid line).}
\label{fig:Figure2}
\end{figure*} 
\indent For infrasound and acoustic measurements, a sine wave voltage signal at frequency 1 Hz-20 kHz was generated by Room EQ Wizard (REW) PC software connected to RME Fireface UFX sound card. The AC voltage is applied in series to a DC bias at the electrodes connected to the graphene aerogel and the generated sound at frequency 1 Hz-20 kHz was acquired by a calibrated microphone (Earthworks M50) connected to the sound card and placed in front of the sample at a fixed distance of $r_{0}=1$ cm for near-field measurements and $r_{0}=1$ m for far-field measurements. Gated measurements \cite{Everest2001} were carried out in order to record the sample sound frequency response without any sound reflections, as in anechoic chamber.\\
\indent For ultrasound measurements, a liquid medium (deionized water 18 M$\Omega$cm) was employed in order to avoid ultrasound absorption by air. A sine wave voltage signal at frequency 20 kHz-20 MHz was generated by a GFG-8210 function generator. The AC voltage was applied in series to a DC voltage at the electrodes connected to the graphene aerogel immersed in a beaker filled with water and the generated sound at frequency 20 kHz-20 MHz was acquired by a calibrated 0.2 mm needle hydrophone (Precision Acoustics) connected to a pre-amplifier, a DC coupler, and an oscilloscope. Since graphene aerogels are superhydrophobic \cite{DeNicola2020a}, the samples do not absorb the water when immersed, therefore their mass density does not increase.\\
\indent Directivity patterns were measured by a goniometric stage with the sample fixed in the center and the microphone rotating around the sample.\\
\indent The sample dynamic range acquired as a function of the sound frequency, was divided by the background noise of the room and smoothed at 1/3 per octave. Therefore, here the SPL corresponds to the signal-to-noise ratio.
\section{Thermal characterization}
\label{sec:thermal}
\indent The graphene aerogel thermal properties were investigated by a Fluke Ti20 thermal camera [Figure \ref{fig:Figure2} (a, b)]. A sine wave voltage signal $V(t)=V_{0}\sin{(2\pi ft)}$ with peak amplitude $V_{0}=1$ V, frequency $f=1$ kHz, and power $Q=V^{2}(t)/Z\equiv V^{2}_{0}[1+\cos{(4\pi ft)}]/2Z$, being $Z=380$ $\Omega$ the modulus of the electrical impedance of the graphene aerogel, was generated by a GFG-8210 function generator. The AC voltage was applied in series to a $V_{DC}=20V_{0}$ DC bias at the electrodes connected to the graphene aerogel. In this way, the first harmonic sound generation is recovered by heterodyning \cite{Heath2017} with an intensity 80 times higher than the second harmonic, which means a total harmonic distortion \cite{Everest2001} about 1\%. However, we could not observe such a small harmonic distortion. Therefore, the power reads $Q\equiv[(V^{2}_{0}+2V^{2}_{DC})+4V_{0}V_{DC}\cos{(2\pi ft)}+V_{0}^{2}\cos{(4\pi ft)}]/2Z=1$ W. It is worth noting that in the thermoacoustic effect only the oscillating component $Q_{0}=[4V_{0}V_{DC}\cos{(2\pi ft)}+V_{0}^{2}\cos{(4\pi ft)}]/2Z$ of the electrical heat $Q$ contributes to the sound pressure generation.\\
\indent Heat is provided to the graphene aerogel by Joule effect, by two electrodes placed at the lateral surface of the cylindrical sample. At first approximation, heat is dissipated into the aerogel by conduction and in the environment by natural convection and radiation. In particular, by solving the Fourier law of heat in cylindrical coordinates, we obtained that in the cylindrical sample the heat dissipated by conduction is $Q_{cond}=4\pi Lk_{s}\left(1-r^{\prime2}/r^{2}\right)^{-1}\left(T\left(r^{\prime}\right)-T_{s}\right)$, with $k_{s}$ the aerogel thermal conductivity, $T\left(r^{\prime}\right)$ the radial temperature in the sample with radius $r=2$ cm and $T_{s}$ the temperature on the sample surface. At the aerogel surface the heat dissipated by radiation equals the heat dissipated by convection $Q_{rad}=Q_{conv}$, where $Q_{rad}=\epsilon\sigma S(T_{s}^{4}-T_{g}^{4})$, with $\epsilon=0.99$ the aerogel emissivity, $\sigma=5.67\cdot10^{-8}$ W/m$^{2}$K$^{4}$ the Stefan-Boltzmann constant, $S=81.6$ cm$^{2}$ the aerogel lateral surface, and $T_{g}$ the temperature of the medium; and $Q_{cov}=hS(T_{s}-T_{g})$, with $h$ the natural convection coefficient. Therefore, $h=\epsilon\sigma(T_{s}^{4}-T_{g}^{4})/\left(T_{s}-T_{g}\right)$, which we estimated for air $h\approx6$ W/m$^{2}$K, while for water $h\approx200$ W/m$^{2}$K \cite{Callen1985}.\\
\indent The heat-up phase was recorded in order to obtain the heat capacity $C_{s}$ of the samples (Figure \ref{fig:Figure2}c). Since the Debye temperature of carbon allotropes is $\Theta_{D}\approx2100$ K \cite{Pop2012}, the heat capacity is not stationary in the range of temperature considered. Therefore, since the steady state is the one that contributes to the sound generation, we took into account the value of heat capacity $C_{s}=10^{-1}\pm10^{-2}$ J/K obtained by fitting data by the law
\begin{equation}
T\left(t\right)=T_{MAX}+(T_{MIN}-T_{MAX})Exp\left[\left(t_{0}-t\right)/\tau\right]
\end{equation}
where $\tau=C_{s}(T_{MAX}-T_{MIN})/Q$, $t_{0}$ is the initial time and $T_{MAX, MIN}$ the sample maximum and minimum temperature measured, respectively.\\
\indent Thermal conductivity $k_{s}$ of the samples was derived by a thermal profile image of the heated sample base surface at the steady state (Figure \ref{fig:Figure2}d), by fitting data with the expression
\begin{equation}
T\left(r^{\prime}\right)=\frac{Q}{4\pi Lk_{s}}\left(1-\frac{r^{\prime2}}{r^{2}}\right)+T_{s},
\end{equation}
The fit returned $k_{s}=10^{-1}\pm10^{-2}$ W/mK.\\
\indent Furthermore, since the Biot number $Bi\equiv hV/(Sk_{s})=0.6>0.1$, with $V=81.6$ cm$^{3}$ the aerogel volume, it suggests a non-negligible internal thermal resistance of the graphene aerogel.\\
\indent Thermal diffusivity $\alpha_{s}$ of the samples was studied by heating the samples on the base surface and recording the heat-up phase at the orthogonal lateral surface (Figure \ref{fig:Figure2}e). Data were fit by the law \cite{Parker1961}
\begin{equation}
T\left(t\right)=T_{MAX}\left[1+2\sum^{\infty}_{n=1}\left(-1\right)^{n}Exp\left(\frac{-n^{2}\pi^{2}\alpha_{s} t}{L^{2}}\right)\right],
\end{equation}
and we obtained $\alpha_{s}=10^{-4}\pm10^{-5}$ m$^{2}$/s.\\
\indent Therefore, the minimum effusivity achieved for the samples is $e_{s}\equiv\sqrt{k_{s}\rho_{s}C_{p,s}}=10.95\pm0.03$ Ws$^{1/2}$/m$^{2}$K, which is larger than the value reported for air (6 Ws$^{1/2}$/m$^{2}$K)\cite{Callen1985} but is lower than the value reported for water (1588 Ws$^{1/2}$/m$^{2}$K)\cite{Callen1985}.
\subsection{Electrical characterization}
\label{sec:impedence}
\begin{figure}[hb]
\centering
\includegraphics[width=8cm, keepaspectratio]{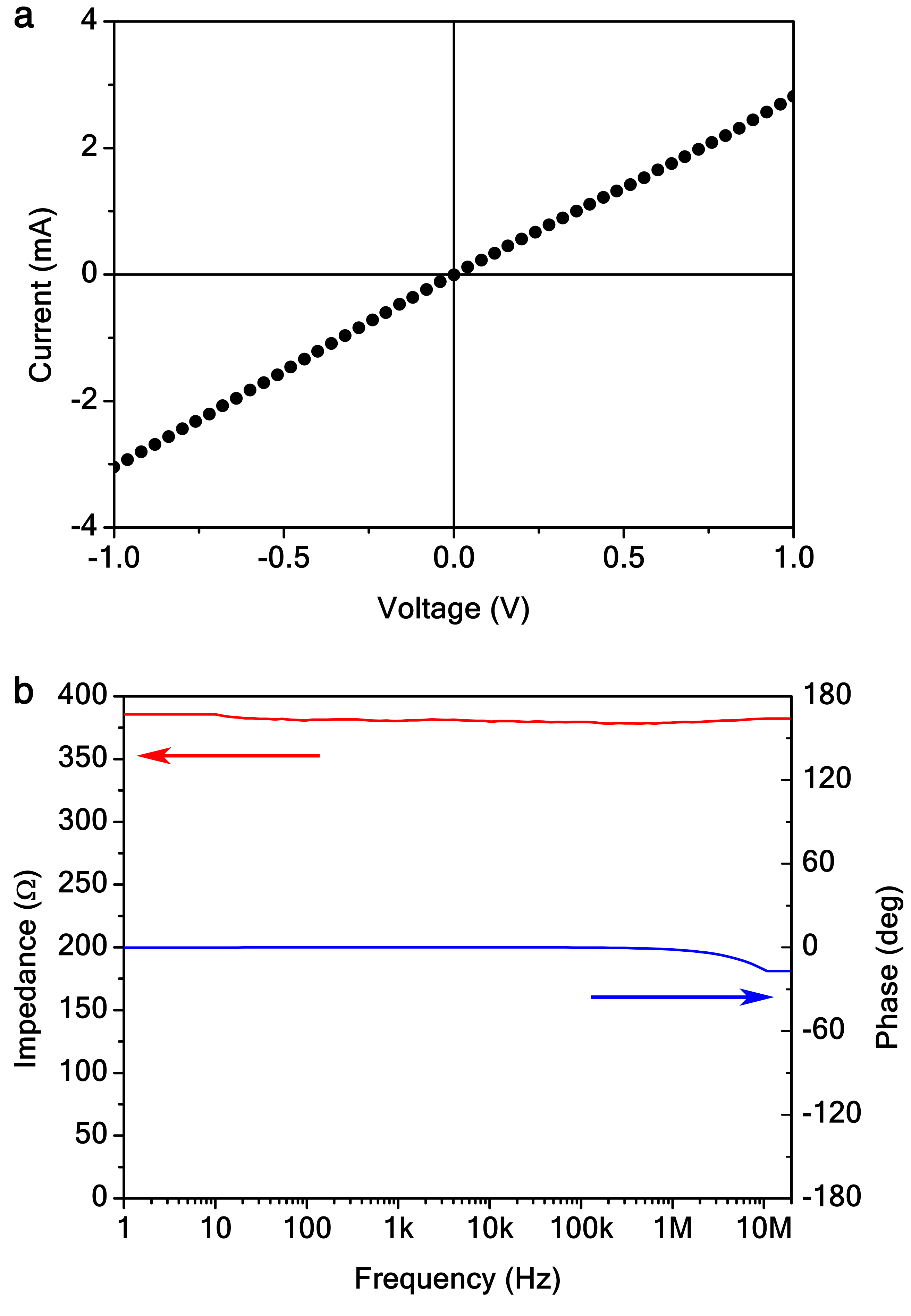}
\caption{\textbf{a}, Electrical current as a function of applied voltage to a graphene aerogel. \textbf{b}, Modulus and phase of the electrical impedance a graphene aerogel as a function of frequency.}
\label{fig:Figure3}
\end{figure}

\indent Current-voltage curves (Figure \ref{fig:Figure3}a) were measured by a Keithley 2612b digital sourcemeter unit. Impedance measurements (Figure \ref{fig:Figure3}b) were carried out by an Agilent HP4192A impedance analyzer.
\begin{figure*}[ht]
\centering
\includegraphics[width=1\textwidth]{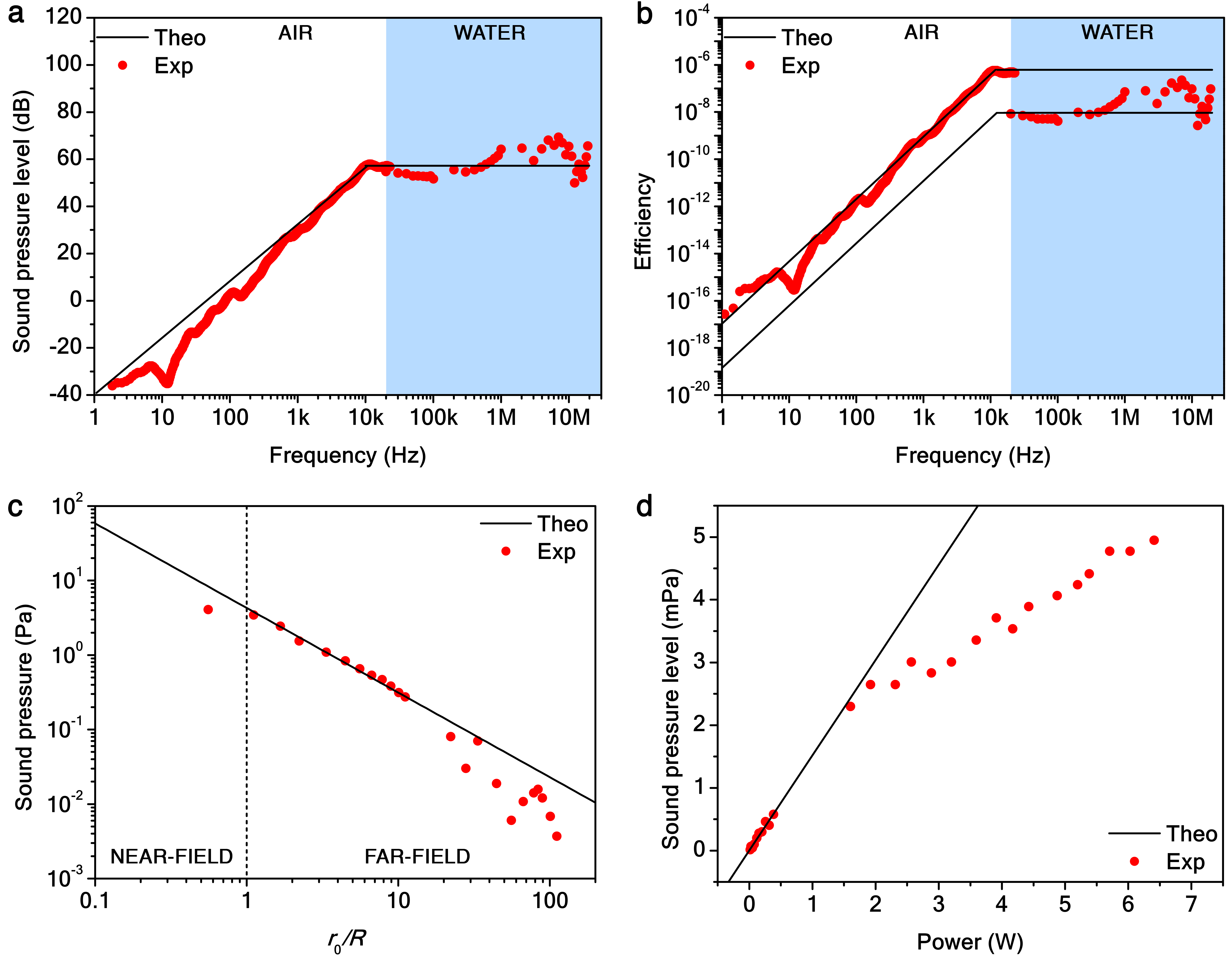}
\caption{Thermoacoustic effect in graphene aerogels. \textbf{a}, Unweighted SPL frequency response at 1 W/1 m for a graphene aerogel (red solid dots). The black solid curve represents the limiting analytical TA model with no free parameters. \textbf{b}, Unweighted TA efficiency frequency response at 1 W/1 m for the aerogel (red solid dots). The black solid curve represents the TA model. \textbf{c}, Sound pressure in air at 20 kHz/1 W  as a function of the distance from the graphene aerogel sound-emitting surface in log-log scale (red solid dots). The black curve represents the TA model $p_{rms}\propto R/r_{0}$, which accounts for the far-field regime. The boundary between the near-field and the far-field regime is set by the Rayleigh criterion for $r_{0}/R=1$. \textbf{d}, Sound pressure in air at 20 kHz/1 m as a function of the electrical power amplitude (red solid dots). The black curve represents the TA model $p_{rms}\propto Q_{0}$ in the linear regime.}
\label{fig:Figure4}
\end{figure*}
\section{Results and discussion}
\label{sec:results}
\indent In Figure \ref{fig:Figure4}a, the frequency response of the sound pressure level \cite{Everest2001} (SPL) $L_{p}=20\log_{10}{(p_{rms}/p_{0})}$, with $p_{0}=20$ $\mu$Pa the RMS pressure of the minimum audible threshold, at the input power of 1 W and rescaled at 1 m distance from the source in air and water, according to the Audio Engineering Society standard for acoustic measurements (AES02-1984-r2003), for a graphene aerogels with mass density $\rho_{s}=6.55\pm0.01$ is shown. It is worth noting that the SPL curves measured in air and water overlap each other, as the product of the specific heat and the effusivity term is similar for the two media. Furthermore, below 200 Hz, where the modal resonances due to the finite size of the sound-proof room decrease the signal-to-noise ratio, experimental data depart from the TA analytical model with no free parameters in Equation \ref{eq:TA}. Also, unlike conventional acoustic transducers\cite{Beranek1993}, the SPL frequency response of graphene aerogels is not dependent on the electrical impedance of the load as the latter is constant up to the range of frequency investigated (see Experimental section). Moreover, owing to the far-field directivity $\mathcal{D}(\theta,\phi)\propto f^{-1}$ in the ultrasound range, the SPL saturates above 20 kHz at about 60 dB, corresponding to a sensitivity \cite{Everest2001} (i.e., the SPL relative to the maximum audible threshold of 120 dB) of -60 dB that is well above the minimum audible threshold (-120 dB) \cite{Everest2001}, and it is the highest value reported for any TA devices in the ultrasound range \cite{Tian2014a,Aliev2010}. For instance, at 20 kHz/1 W/1 cm or 20 kHz/1 kW/1 m the sensitivity is 0 dB. The graphene aerogel sensitivity may be increased further by increasing its surface area. In this way, the sound emitted by every point on the aerogel heated surface by Joule effect coherently add up to increase the SPL. By comparison, commercial earphones have a similar sensitivity in air as graphene aerogels at 1 W/1 m/1 kHz, due to their similar surface area \cite{DeNicola2019}. On the other hand, graphene aerogels act as a highly damped transducer in the ultrasound range \cite{Hedrick2005}. Therefore, graphene aerogels could be successfully used as TA broadband transducers.\\
\indent According to the TA model\cite{Hu2010}, the TA efficiency (i.e., the ratio between the output acoustic power $P_{ac}$ and input electrical power $Q_{0}$) reads
\begin{equation}
\eta=\frac{I}{Q_{0}}r_{0}^{2}\int_{0}^{2\pi}\int_{0}^{\pi/2}\mathcal{D}^{2}(\theta,\phi)\sin{\theta}d\theta d\phi,
\end{equation}
where $I=p_{rms}^{2}/\rho_{g}v_{g}$ is the sound intensity, with $\rho_{g}$ the mass density the medium. In Figure \ref{fig:Figure4}b, the TA efficiency at 1 W/1 m measured in air and water for the graphene aerogel along the theoretical curves are reported. The TA efficiency saturates in the ultrasound range at about $10^{-7}$ in air and $10^{-9}$ in water due to the directivity integral that is proportional to $f^{-2}$. On the other hand, the thermodynamic efficiency of the TA process $\eta^{\prime}\equiv1-T_{c}/T_{h}=30\%$ is given by a Carnot cycle \cite{Vesterinen2010,Aliev2015} between two heat reservoirs of temperature $T_{c}=298$ K and $T_{h}=430$ K. Evidently 70\% of the heat provided to the aerogel is lost in the environment (20\% radiatively, 20\% convectively, and 30\% owing to the sample internal thermal resistance $R_{i}\equiv1/h_{i}S=433$ K/W) without generation of sound.\\
\indent In Figure \ref{fig:Figure4}c, we show that the experimental sound pressure at 20 kHz/1 W departs from the far-field TA model for $r_{0}/R<1$, in agreement with the Rayleigh criterion that defines the boundary between near-field and far-field \cite{Blackstock2000}. In addition, Figure \ref{fig:Figure4}d illustrates that for values of electrical power $Q_{0}>1.5$ W the experimental sound pressure at 20 kHz/1 m departs from the linear TA model, partially due to the air nonlinear thermal conductivity $k_{g}\propto\sqrt{T}$ \cite{Aliev2015} and the graphene aerogel nonlinear specific heat below the value of Debye temperature $\Theta_{D}\approx2100$ K for carbon allotropes \cite{Pop2012} (see Experimental section).\\
\begin{figure}[ht]
\centering
\includegraphics[width=0.4\textwidth]{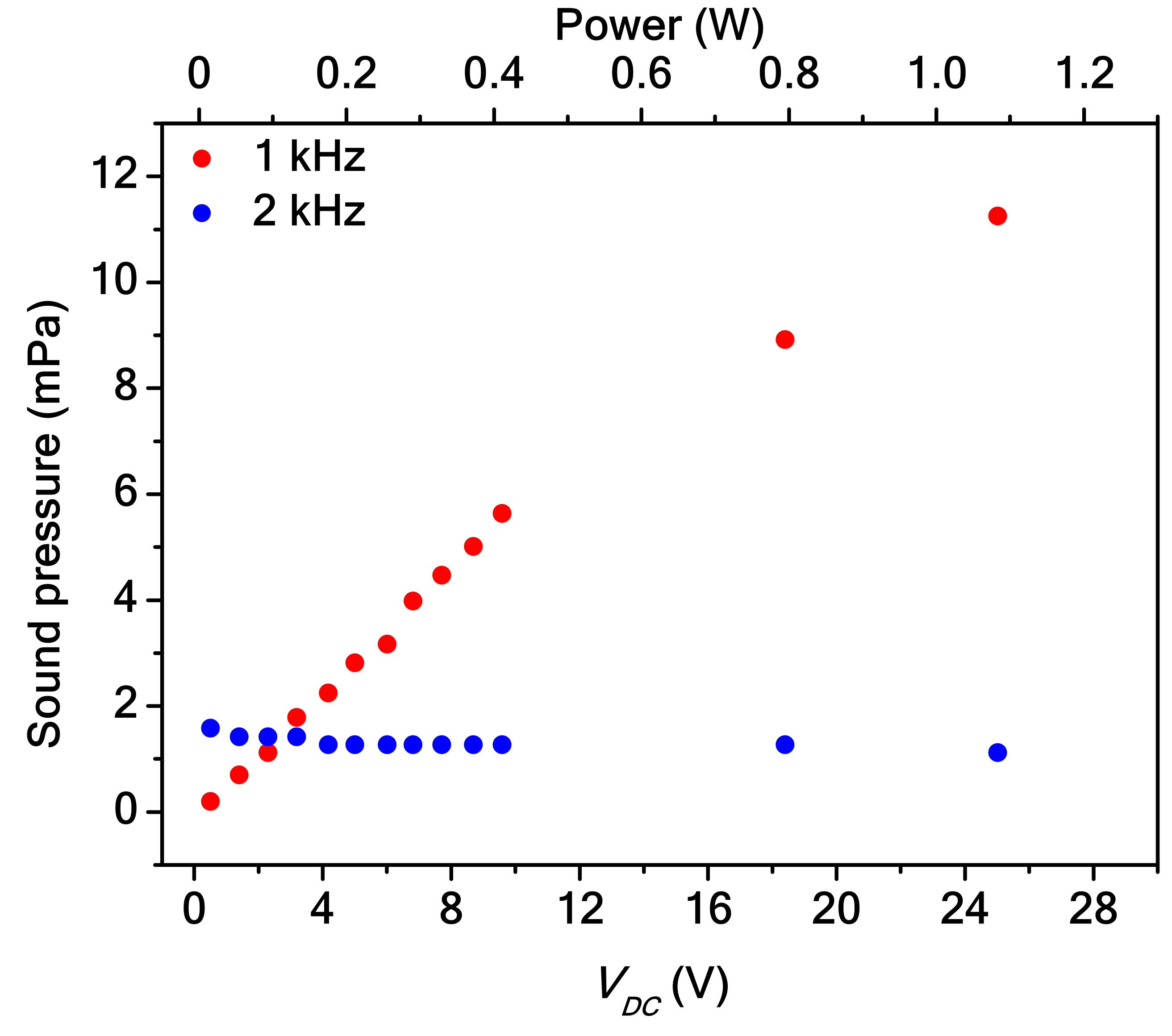}
\caption{Sound pressure recorded in air at 1 m distance as a function of DC voltage $V_{DC}$ applied to the sample with the AC voltage set to $V_{0}=9$ V, for the first (1 kHz) and second (2 kHz) harmonic generated by an input electrical signal of 1 kHz.}
\label{fig:Figure5}
\end{figure}
\begin{figure}[hb!]
\centering
\includegraphics[width=0.37\textwidth]{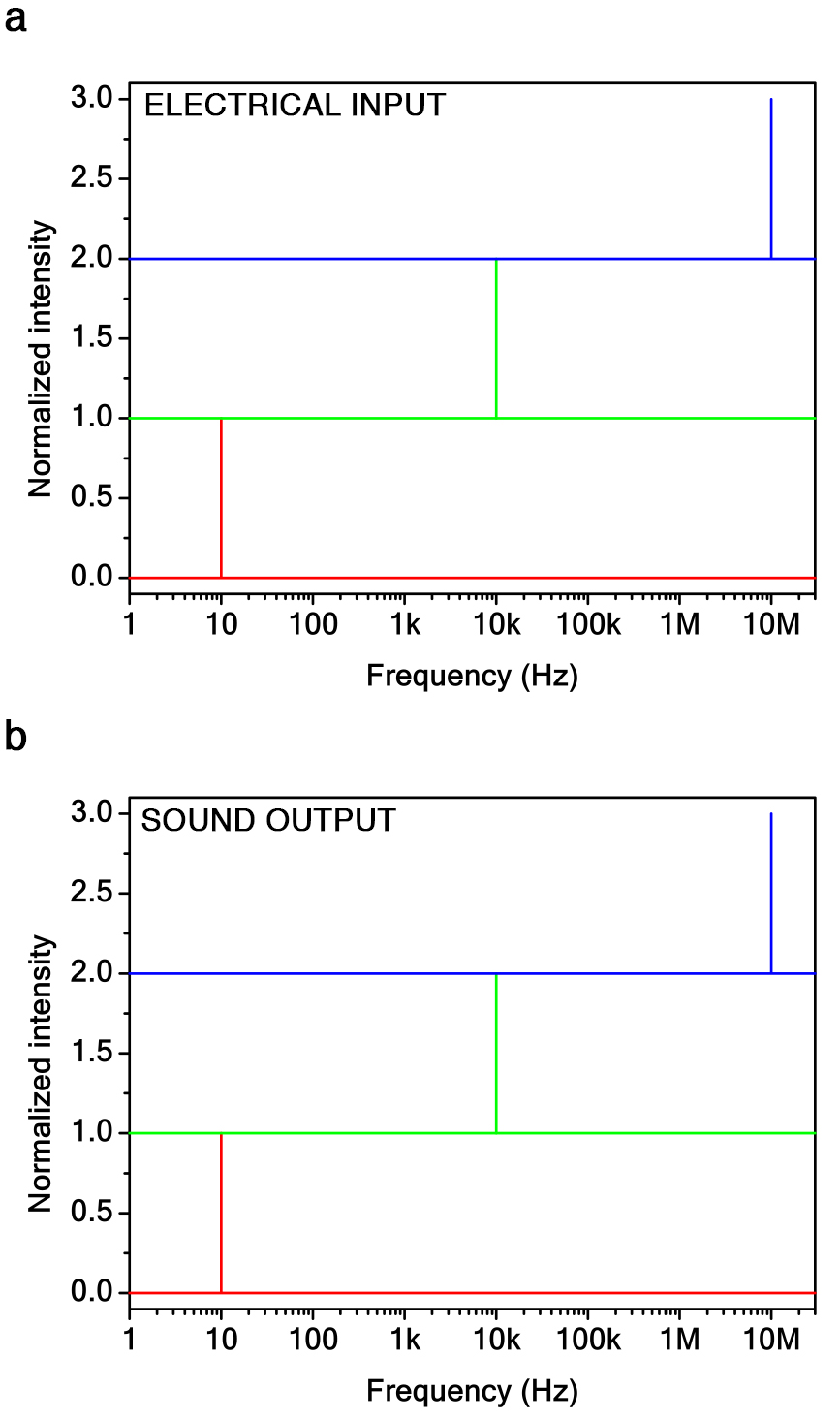}
\caption{Harmonic analysis of TA sound emission in graphene aerogels. Fast Fourier Transform of the electrical input (\textbf{a}) and sound output (\textbf{b}) signals at 10 Hz, 10 kHz, and 10 MHz. The input and output signals are undistorted from the infrasound to the audible, and up to ultrasound range investigated.}
\label{fig:Figure6}
\end{figure}
\begin{figure}[ht!]
\centering
\includegraphics[width=0.405\textwidth]{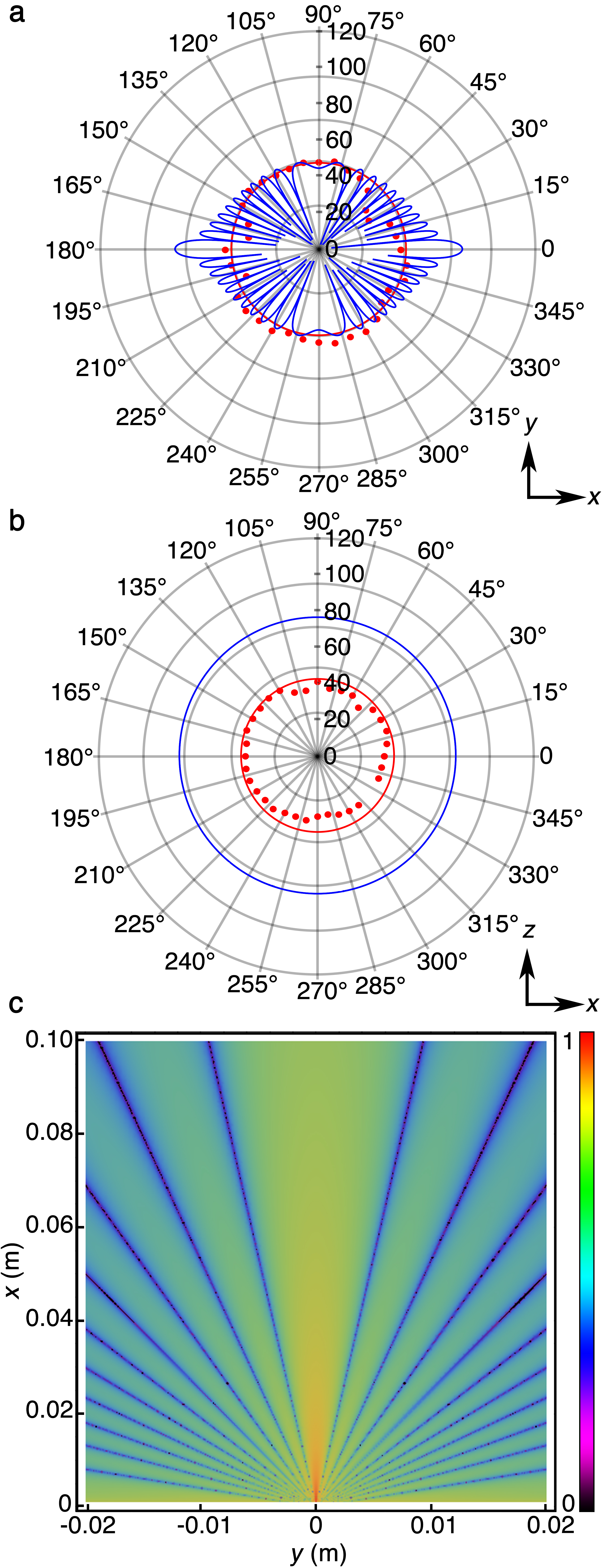}
\caption{Thermoacoustic directivity in graphene aerogel transducers. Experimental sound pressure level at 20 kHz/1 W/1 m as a function of the angle in the azimuthal (\textbf{a}) and polar (\textbf{b}) planes (red dots). The TA model is reported at 20 kHz (red solid curve) and 1 MHz (blue solid curve). The directivity indexes \cite{Beranek1993} are DI(20 kHz)=11 dB and DI(1 MHz)=40 dB. \textbf{c}, Calculated sound pressure level in space at 1 MHz/1 W/1 m in the azimuthal plane by the TA model. The sound pressure level is normalized to its maximum value.}
\label{fig:Figure7}
\end{figure}
\indent Furthermore, in Figure \ref{fig:Figure5} we confirmed that by heterodyning only the sound pressure of the first harmonic is linearly dependent on $V_{DC}$, at low power. The nonlinear behavior observed at higher power is due to the nonlinear specific heat of the graphene aerogel. Therefore, it is possible to amplify the SPL of a graphene aerogel by applying a small DC voltage in series to the AC voltage applied to the sample.\\
\indent In Figure \ref{fig:Figure6} (a, b), the frequency response of the input (electrical) and output (sound) signals are respectively reported in the infrasound (10 Hz), audible (10 kHz), and ultrasound (10 MHz) range to evaluate the harmonic distortion of the sound generated by the graphene aerogel. We noted no harmonic distortion in graphene aerogels over the range investigated. It is worth noting that in our experiments the first harmonic generation is obtained by heterodyning \cite{Heath2017}, as a DC voltage is applied in series to the AC voltage to the samples. By comparison, commercial hi-fi loudspeakers and earphones typically have a total harmonic distortion (THD) about 1\% \cite{DeNicola2019}.\\
\indent The theoretical and experimental directivity pattern \cite{Beranek1993} of the sound emitted at 1 W/1 m from the graphene aerogel are reported in the azimuthal (Figure \ref{fig:Figure7}a) and polar planes (Figure \ref{fig:Figure7}b). The samples behave as an omnidirectional point source up to 20 kHz, beyond which they present a multi-polar pattern, as shown for instance at 1 MHz. Since there are no electro-mechanically moving parts in the emitter, the sound is equally generated with no destructive interference from both the sides of the illuminated spot of the aerogel \cite{DeNicola2019}, which acts as a diaphragm of thermal thickness $\mu\equiv\sqrt{\alpha_{s}/\pi f}=0.1$ $\mu$m-1 mm, in the range of frequency studied. Hence, no bass reflex technique \cite{Beranek1993} is needed in TA transducers. Moreover, Figure \ref{fig:Figure7}c depicts the SPL calculated in space at 1 MHz, in the azimuthal plane. Interestingly, the sound is suppressed at given angles corresponding to the nodes in the directivity pattern of Figure \ref{fig:Figure7}a.
\section{Conclusions}
\label{sec:conclusions}
\indent Graphene aerogels can be effectively employed as ultrabroadband TA transducers with omnidirectional emission from 1 Hz to 20 kHz and with no harmonic distortion from 1 Hz to 20 MHz, exploiting a single, non-mechanically vibrating emitter. For instance, such devices may be improved by using thicker graphene aerogels with lower electrical impedance in order to require lower voltage to operate.\\ 
\begin{acknowledgments}
The authors acknowledge that this project has received funding from the European Union's Horizon 2020 research and innovation programme under grant agreement No. 785219 - GrapheneCore2. The authors also acknowledge that this project has received the financial support of the Bilateral Cooperation Agreement between Italy and China of the Italian Ministry of Foreign Affairs and of the International Cooperation (MAECI) and the National Natural Science Foundation of China (NSFC), in the framework of the project of major relevance 3-Dimensional Graphene: Applications in Catalysis, Photoacoustics and Plasmonics.\\
The authors declare no competing financial interests.
\end{acknowledgments}
\end{document}